\documentclass[reprint,amsmath,amssymb,aps,prb,letterpaper,showpacs]{revtex4-1}

 \usepackage{graphicx}
\usepackage{dcolumn}
\usepackage{bm}
\usepackage{lipsum}  
\usepackage{hyperref} 
\usepackage[mathlines]{lineno}
\usepackage[top=2.8cm, bottom=2.8cm, left=2.4cm, right=2.4cm]{geometry}

\begin{document}

\preprint{APS/123-QED}



\title{Phase Diagrams Construction Using Mean-Field Renormalization and Neural Network Fitting}

\author{Esteban Bedoya-Rodriguez}
\email{juan.esteban.bedoya@correounivalle.edu.co}
\affiliation{Department of Physics, Universidad del Valle, Colombia}

\author{Leon Escobar-Diaz}%
\email{leon.escobar@correounivalle.edu.co}
\affiliation{Department of Mathematics, Universidad del Valle, Colombia}

\author{Sebastian Trujillo-Hernandez}%
\email{juan.sebastian.trujillo@correounivalle.edu.co}
\affiliation{Department of Physics, Universidad del Valle, Colombia}

\begin{abstract}

Employing the mean-field renormalization group (MFRG) method, we analyzed the $Fe_p Mn_{0.6-p}Al_{0.4}$ and $Fe_{p}Al_{1-p}$  alloys, incorporating second-neighbor interactions in the latter for the first time within this framework. Our analysis utilized neural network fitting, yielding promising results in both the phase diagram adjustments and the estimation of bond energies.

\end{abstract}


\maketitle

\section{INTRODUCTION}

Magnetism in materials has long been one of the most thoroughly studied areas of physics, encompassing both experimental and theoretical perspectives. Experimentally, the fields of physical metallurgy and the investigation of materials with exotic magnetic properties have been particularly prominent. Theoretical studies have evolved considerably, progressing from a classical understanding of magnetism—initially attributed to electron orbitals—to a quantum mechanical view centered on spin. These advancements have led to a deeper understanding of various magnetic behaviors, such as paramagnetism (explained by Weiss and Langevin), ferromagnetism (described through Weiss's molecular field, now interpreted as spin interactions), and antiferromagnetism (developed by Néel using Weiss's molecular field) \cite{cullity2011introduction,kittel2018introduction}. The application of quantum mechanics has further refined these concepts, translating them into spin-based models, which not only provide more accurate interpretations of these magnetic phases but also explain phenomena like spin glass and reentrant spin glass phases.

Within this quantum framework, the study of magnetism is best approached through spin dynamics, using tools from statistical mechanics. In this work, we employ the mean-field renormalization group (MFRG) method, which has shown promising results in previous studies \cite{Perez1,Enlace_pares_segunda,Blume_Capel,FeNiMn,Perez2,Blume-Capel2,Blume-Capel0}. Specifically, we revisit the diluted and random bond model introduced by Osorio et al. \cite{Enlace_pares_primera} and recently applied by Pérez et al. \cite{rodriguez2023} in order to study $Fe_{p}Al_{1-p}$ and $Fe_{p}Mn_{0.6-p}Al_{0.4}$  systems, which are of particular interest due to their unique magnetic properties. It is well known that the first exhibits significant second-neighbor interactions, while the later system features spins with competing bond energies, leading to complex phases such as spin glass and reentrant spin glass.

\begin{figure}[htbp]
    \centering
    \includegraphics[width=1\linewidth]{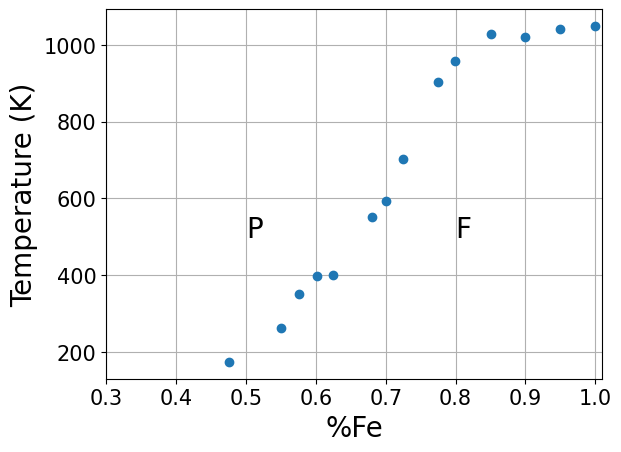}
    \caption{Experimental data from the magnetic phase diagram of the $Fe_{p}Al_{1-p}$ system disordered, where changes in the critical temperature of the system with respect to the iron concentration are observed. The data were obtained from reference \cite{datosFeAl,datosFeAl2,adatosFeAl3,adatosFeAl4}.}
    \label{FeMnAl_exp_dat}
\end{figure}

\begin{figure}[htbp]
    \centering
    \includegraphics[width=1\linewidth]{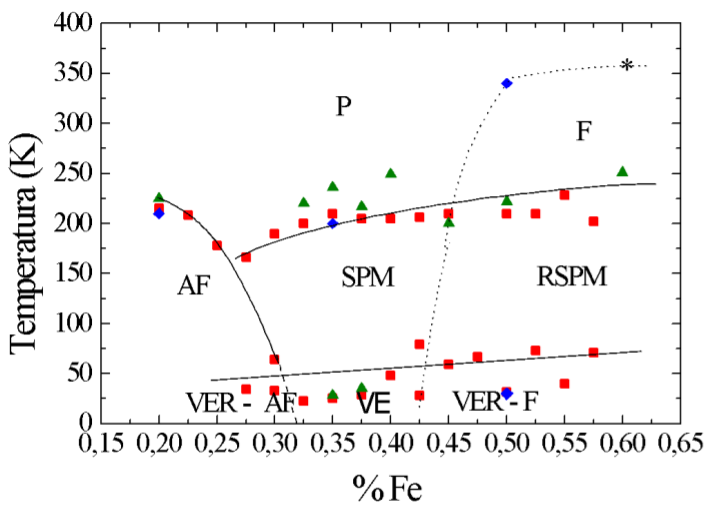}
    \caption{Experimental data from the magnetic phase diagram of the $Fe_{p}Mn_{0.6-p}Al_{0.4}$ system disordered, where changes in the critical temperature of the system with respect to the iron concentration are observed. The data were obtained from reference \cite{datosFeMnAl}.}
    \label{FeAl_exp_dat}
\end{figure}

Our study is based on the experimental data presented in Figures \ref{FeMnAl_exp_dat} and \ref{FeAl_exp_dat}. To fit this data to the model derived from the MFRG method, we employed an uninformed neural network approach. Specifically, we used this method to solve the equations that describe the behavior of temperature and bond energy at the desired phase transition points, by modeling these equations as a neural network dependent on certain parameters to be determined. Although this approach is similar to methods used to solve differential equations, it is versatile enough to be applied, in principle, to equations of any nature with an appropriate choice of layers, boundary conditions, and activation functions. For a recent review of this topic, see \cite{blechschmidt2021three}. In our particular case, the equations we aim to solve are algebraic in nature.

This work is structured as follows: In Section \ref{sec:theoretical_model}, we provide a concise overview of the MFRG method. Later, in Section \ref{sec:recurrence_relations}, using the previous expressions, we derive the algebraic equations that describe various magnetic material systems, such as \( Fe_pMn_{0.6-p}Al_{0.4} \) and \( Fe_p Al_{1-p} \), while considering interactions with both first and second neighbors. Section \ref{sec:neural_network_approach} outlines the application of an unconstrained neural network approach to solve these algebraic equations for alloy systems. In Section \ref{sec:numerical_results}, we present and discuss the numerical results, and finally, in Section \ref{discussion}, we offer our conclusions of this work.

\section{THE MFRG MODEL}\label{sec:theoretical_model}
The concept of the MFRG method was initially introduced by J. O. Indekeu, A. Maritan, and A. L. Stella in 1982 \cite{primero}. It builds upon the classical framework of renormalization groups proposed by Kadanoff in 1966 \cite{kandanoff}, incorporating mean-field concepts. The approach involves analyzing a large system of spins by dividing it into two groups of sizes $N$ and $N'$, respectively. Given that all state functions are homogeneous, it follows that

 \begin{equation}
     f_N(J,b)=\alpha f_{N'}(J',b') \label{fun_estado}.
 \end{equation}
 
Where \( f_N \) denotes the state function, \( J \) represents the coupling energy between spins, and \( b \) is the mean field experienced by the cluster with \( N \) spins. Similarly, analogous definitions apply for \( f_{N'} \), \( J' \), and \( b' \) in the cluster with \( N' \) spins. Additionally, since magnetization is a state function of the system and the mean field is proportional to it, this field can be rescaled in a similar way, yielding \( b = \alpha b \). Thus, when analyzing the system’s behavior at critical transition points, it is reasonable to assume that \( b \rightarrow 0 \). By substituting this condition into expression \ref{fun_estado}, we can apply a Taylor expansion of the state functions in \( b \) around \( 0 \) to obtain 
 
 \begin{equation}
     \Gamma(K,b,N)=\alpha \Gamma'(K',b',N')\label{ecuarenorma_previa},
 \end{equation}
 where 
 \begin{equation*}
     \Gamma(K,b,N)=f_N(K,b)\biggl|_{b\rightarrow \ 0}+\frac{\partial f_{N}(K,b) }{\partial b} \biggl|_{b\rightarrow \ 0}b+...,
 \end{equation*}
 and similarly for $\Gamma'(K',b',N)$.
 Considering a first-order approximation in the last expressions and using the condition $b = \alpha b'$ in equation \ref{ecuarenorma_previa}, we obtain that
\begin{equation}
    \frac{\partial f_{N}(K,b) }{\partial b} \biggl|_{b\rightarrow \ 0}=\frac{\partial f_{N'}(K',b') }{\partial b'}\biggl|_{b'\rightarrow \ 0} \label{EcuaRenorma},
\end{equation}
where \( K \) and \( K' \) are rescaled forms of the coupling energy, expressed as \( K =  J /( k_B T)  \) and \( K' = J'/(k_B T) \) respectively with $k_B$ being the Boltzmann constant \cite{beale1996statistical}. Following Indekeu \cite{primero}, this expression will serve as a recurrence relation to analyze the system's behavior at phase transition points (critical points) or fixed points (\( K = K' = K_c \)). However, since it does not explicitly differentiate between single-type and multi-type spin systems, this work will utilize a configurational average of the coupling energy, as proposed by M. Droz, A. Maritan, and A. L. Stella in 1982 \cite{segundo}. This approach is chosen because the coupling energy is the only variable in equation \ref{EcuaRenorma} that conveys information about the type of spin and its interactions, thereby linking the MFRG method to a system of disordered spins with random coupling, besides, this is important because the studied systems are disordered as a consequence of the preparation method, as well as the thermal average. The configurational average will be applied to the state functions \( f_N \) already used in \ref{EcuaRenorma} as follows

 \begin{equation}
    \overline{f_{N}}=\int f_{N}(K_{i,j}, b) \prod_{i,j}^{N} P(K_{i,j})dK_{i,j},\label{prome}
\end{equation}
 where \( K_{i,j} \) is the spin binding energy between two spin particles located at positions \( i \) and \( j \), respectively, within the group of \( N \) spins. The term \( P(K_{i,j}) \) represents the probability distribution of spin couplings, which can be written as
\begin{equation*}
    P(K_{i,j})=\sum^{n}_{l}p_l\delta(K_{i,j}-\hat K_l),
\end{equation*}
where \( n \) is the number of different bonds in the system,   
\( \hat K_l \) is the binding energy of the \( l \)-th type due to the different interactions between the alloy elements, and \( p_l \) is the probability of finding this type of bond, which is expressed as a product of concentrations in this case. Finally, note that in the above expression $\delta$ denotes the Dirac function.

\section{RECURRENCE RELATIONS}\label{sec:recurrence_relations}
 Since we aim to investigate systems with spin interactions, we will adopt the approach of the well-known Ising model \cite{Perez1}. Thus, let us assume a grid of point particles with spins that interact with their nearest neighbors according to a specific set of rules, where the dimension of the system is determined by the coordination number, or the number of nearest neighbors. Clearly, the energy of the system is given by the following expression
\begin{equation}
    \mathcal{H}=-\sum_{i,j}^{N} J_{i,j} \sigma_{i}\sigma_{j} \label{Ising},
\end{equation}
where $\sigma_{i}$ and $\sigma_{j}$ represent the spin values at sites $i$ and $j$ respectively, and $J_{i,j}$ will be a coupling energy  that  determines the strength of the interaction between neighboring spins.
Note that if $J > 0$, the interaction is ferromagnetic, meaning that neighboring spins tend to align in the same direction. In the contrary case, if $J < 0$, the interaction is antiferromagnetic, causing neighboring spins to prefer opposite directions. Finally, note that  
 the sum is taken over allpairs of neighboring spins $i, j $.

\subsection{Case of nearest neighbors} 
Following the works \cite{primero,segundo,Perez1}, we will consider equation \ref{EcuaRenorma} for two clusters, with \( N = 1 \) and \( N' = 2 \). Furthermore, we will utilize the magnetization per spin as the state function or order parameter for the ferromagnetic and antiferromagnetic transitions. Starting from the Ising Hamiltonian in equation \ref{Ising}, we can derive two rescaled Hamiltonians that represent the systems with \( N = 1 \) and \( N = 2 \) spins, respectively, by applying the mean field method, as follows
\begin{align*}
    \mathcal{H}_1=&-\sum_{i}^{z} K'_{1,i}b'_{1,i}\sigma_{1}  , \\
    \mathcal{H}_2=&-\sum_{i}^{z-1} K_{1,i}b_{1,i}\sigma_{1}-K_{1,2}\sigma_{1}\sigma_{2} -\sum_{i}^{z-1} K_{i,2}b_{2,i}\sigma_{2},
\end{align*}
where \( K_{i,j} := J_{i,j}/ (k_B T) \) , and \( z \) is the coordination number, or the number of nearest neighbors. By utilizing these expressions and considering that the mean field is small at the transition points \( b_{i,j} \rightarrow 0 \), we can derive their respective partition functions and, consequently, the spin magnetization \( \langle \sigma_i \rangle \), which are presented, respectively as
\begin{align}
    \langle \sigma_{1}\rangle_1=&\sum_{i}^{z} K_{1,i}'b_{1,i}'  , \label{magne1}\\
    \langle \sigma_{1}\rangle_2=&\sum_{i}^{z-1} K_{1,i}b_{1,i}+\tanh(K_{1,2})\sum_{i}^{z-1} K_{2,i}b_{2,i}  . \label{magne2}
\end{align}
To incorporate the concept of random bonding into the previously obtained magnetizations, it is essential to calculate the configurational average, which as illustrated in equation \ref{prome}, should be computed as
\begin{equation}
    \overline{\langle\sigma_\nu \rangle}=\int \langle\sigma_i\rangle_\nu \prod_{i,j}^{N} P(K_{i,j})dK_{i,j}, \label{sigma_prome}
\end{equation}
for the index $\nu=1,2$. Then, we can represent the probability densities of the \( Fe_pAl_{1-p} \) and \( Fe_pMn_{0.6-p}Al_{0.4} \) systems, respectively, by the following expressions
\begin{align}
    P(K_{i,j})&=p^{2}\delta(K_{i,j}-K_{0}(p))+(1-p^2)\delta(K_{i,j}) \label{modeloFeAl}, \\  \nonumber \\
    P(K_{i,j})&=p^{2}\delta(K_{i,j}-K_1(p))+x^{2}\delta(K_{i,j}-K_{2}(p)) \nonumber\\
    & \quad +2px\delta(K_{i,j}-K_{3}(p))+q(2-q)\delta(K_{i,j}). \nonumber \\ & \label{modeloFeMnAl} 
\end{align} 
 
In the first expression, given by Equation \ref{modeloFeAl}, \( p \) represents the concentration of iron in the system \( Fe_p Al_{1-p} \), while \( 1 - p \) denotes the concentration of aluminum. The first term corresponds to the bonds between iron atoms, represented as \( \text{Fe-Fe} \) bonds, while the second term represents the bonds between iron and aluminum atoms, as well as the bonds between two aluminum atoms, denoted as \( \text{Fe-Al} \) and \( \text{Al-Al} \), respectively.

Similarly, in the second expression, equation \ref{modeloFeMnAl}, which models the \( Fe_pMn_{0.6-p}Al_{0.4} \) system, the variables \( p \), \( x=0.6-p \), and \( q=0.4 \) represent the concentrations of iron, manganese, and aluminum, respectively. The first term of the expression corresponds to the Fe-Fe bonds, the second to manganese-manganese bonds, the third to Fe-Mn bonds, and the final term represents the bonds involving aluminum. In the following, we will analyze each of these two systems separately.

 \subsubsection{\texorpdfstring{$Fe_{p}Mn_{0.6-p}Al_{0.4}$}. system}

Using the probability distribution from equation \ref{modeloFeMnAl}, along with the magnetizations from equations \ref{magne1} and \ref{magne2}, and the configurational average form for the magnetization in equation \ref{sigma_prome}, we can derive the configurational average of the magnetization, as shown as follows
\begin{align*}
    \overline{\langle \sigma_{1}\rangle_1}=&zb \biggl[ \pm p^{2}K'_1(p)\pm x^{2}K'_{2}(p) \pm 2pxK'_{3}(p) \biggl],  \nonumber \\ \nonumber\\
    \overline{\langle \sigma_{1}\rangle_2}=&(z-1)b \ \biggl[ \pm p^{2}K_1(p)\pm x^{2}K_{2}(p) \pm 2pxK_{3}(p) \biggl] \nonumber\\
    &\biggl[1\pm p^{2} \tanh(K_1(p))\pm x^{2} \tanh(K_2(p))\nonumber\\
    &\pm (2px) \tanh(K_3(p))\biggl],
\end{align*}
where we use the notation \( \overline{b_{i,j}} = \pm b \) \cite{primero}, which corresponds to the two types of magnetic phases present in the system: “+” for the ferromagnetic phase and “-” for the antiferromagnetic phase. 

Using equation \ref{EcuaRenorma}, we derive the recurrence equation for the system. By evaluating this at the fixed point \( (K_\mu = K'_{\mu}) \) for $\mu=1,2,3$, we obtain 
\begin{align}
    0&= -\frac{z}{(z-1)}+1\pm p^{2} \tanh\Big(\frac{J_1(p))}{T  k_B}\Big)\nonumber\\
    &\quad\pm x^{2} \tanh\Big(\frac{J_2(p)}{T k_B}\Big)\pm (2px) \tanh\Big(\frac{J_3(p)}{T k_B}\Big)\label{expModeFeMnAl},
\end{align}
where we have used $ K_\mu(p) := J_\mu (p) /  (k_B T)$. This last expression models the behavior of $J$ and $T$ at the phase transition points of ferromagnetic-paramagnetic and antiferromagnetic-paramagnetic types.

As previously mentioned, the system under consideration exhibits various exotic magnetic phases, such as spin glass phases. These phases are characterized by the fact that magnetization cannot serve as the order parameter for their transition to paramagnetism, as the magnetization is zero in both phases. Therefore, the Edwards-Anderson parameter (see for instance \cite{stein2013spin}) is employed as the state function for these types of transitions, which is calculated as \( q = \langle \sigma \rangle^2 \). This order parameter is derived from equations \ref{magne1} and \ref{magne2}, leading to
\begin{align*}
    q_1=&\sum_{i,j}^{z} K_{1,i}'b_{1,i}'K_{1,j}'b_{1,j}' \,\\
    q_2=&\sum_{i,j}^{z-1} K_{1,i}'b_{1,i}'K_{1,j}'b_{1,j}'\nonumber\\
    &+2\tanh(K_{1,2})\sum_{i,j}^{z-1} K_{2,i}b_{2,i}'K_{1,j}'b_{1,j}'\nonumber\\
    &+\tanh^2(K_{1,2})\sum_{i,j}^{z-1} K_{2,i}b_{2,i}'K_{2,j}'b_{2,j}'.
\end{align*}
Thus, using the probability density from equation \ref{modeloFeMnAl} and considering that in these transitions the mean field is null, behaving such that \( \overline{b_{i,j}b_{l,m}} = \delta_{i,l} \delta_{j,m} h \) \cite{segundo,Perez1} with $\delta_{i,l}$ denoting the usal kronecker delta and $h$ is the square of the mean field average, we obtaint their configurational average as
\begin{align*}
    \overline{q_1}=&zh \biggl[  p^{2}K'_1(p)^2+ x^{2}K'_{2}(p)^2 + 2pxK'_{3}(p)^2 \biggl]  \\ \nonumber\\
    \overline{q_2}=&(z-1) h \biggl[ p^{2}K'_1(p)^2+ x^{2}K'_{2}(p)^2 + 2pxK'_{3}(p)^2 \biggl] \nonumber\\
    &  \biggl[1+p^{2}\tanh^2(K_1'(p))+ x^{2}\tanh^2(K'_{2}(p)) \nonumber  \\ 
    &+ 2px\tanh^2(K'_{3}(p))\biggl] .
\end{align*}
Finally, by equating the \( q \) parameters of the spin groups with \( N = 1 \) and \( N = 2 \) as given in the two last expressions, using equation \ref{EcuaRenorma}, and evaluating them at their fixed points , we obtain the expression that models the spin glass transition of the  system \( Fe_pMn_{0.6-p}Al_{0.4} \), which takes the following form
\begin{align}
   0=&-\frac{z}{(z-1)}+1 +p^{2} \tanh^{2}\Big(\frac{J_1(p)}{T  k_B}\Big) \nonumber\\
    &+x^{2} \tanh^{2}\Big(\frac{J_{2}(p)}{T k_B}\Big)+(2px) \tanh^{2}\Big(\frac{J_{3}(p)}{T k_B}\Big)\label{expModeFeMnAlSg},
\end{align}
From this point onward, we will refer to it as the state equation of the system. Furthermore, we will denote the right-hand side of this equation by \( \mathcal{E}(J_1(p), J_2(p), J_3(p)) \), such that we write the above equation as
\begin{equation}\label{eq1}
    \mathcal{E}(J_1(p), J_2(p), J_3(p)) = 0. 
\end{equation}

\subsubsection{\texorpdfstring{\(  Fe_pAl_{1-p}\)}. system}

In a similar manner to the previous system, for the \( Fe_pAl_{1-p} \) case, we can obtain the configurational average of the magnetization by using equations \ref{magne1} and \ref{magne2}, along with the probability distribution provided in equation \ref{modeloFeAl} and following the approach outlined in equation \ref{sigma_prome}. This leads to

\begin{align*}
    \overline{\langle \sigma_{1}\rangle_1}=&zbp^{2}K'_0(p),\\
    \overline{\langle \sigma_{1}\rangle_2}=&(z-1)bp^{2}K_0(p)\biggl[1+ p^{2} \tanh(K_0(p))\biggl].
\end{align*}
which, by applying equation \ref{EcuaRenorma}, allows us to obtain the recurrence equation for the system. Evaluating this at the fixed point \( (K = K') \), we obtain
\begin{align}
  0 = \frac{-z}{(z-1)} + &1+p^{2} \tanh\Big(\frac{J_0(p)}{Tk_B}\Big),\label{expModeFeAl}
\end{align}
where \( K_0 :=J_0 / (k_B T)  \). Similarly as in the previous case, we will denote the right hand side this expression as $\mathcal{E}(J_0(p))$, which leads to
\begin{equation}\label{eq2}
    \mathcal{E}(J_0(p)) = 0. 
\end{equation}

\subsection{Case of second-neighbors}
In the study of the \( Fe_pAl_{1-p} \) system and its spin interactions, it has been observed that second-neighbor interactions of the iron-aluminium-iron type are present. These interactions occur through superexchange and are generally antiferromagnetic in nature \cite{veci2,veci22}. Consequently, it is necessary to incorporate them into the probability density and, more broadly, within the framework of the RGMF method. To do this, one must begin with a Hamiltonian that includes second-neighbor interactions. This is achieved by starting from the Ising model, partitioning it into groups of one and two spins, similar to the treatment used for the first-neighbor Hamiltonian, resulting in
\begin{align}
    \mathcal{H}_1=&-\sum_{i}^{z} K_{1,i}'b_{1,i}'\sigma_{1}-\sum_{i}^{z'} K_{1,i}'^{*}b_{1,i}'\sigma_{1} \ ,\label{h12}\\
    \mathcal{H}_2=&-\sum_{i}^{z-1} K_{1,i}b_{1,i}\sigma_{1}-K_{1,2}\sigma_{1}\sigma_{2} -\sum_{i}^{z-1} K_{i,2}b_{2,i}\sigma_{2}\nonumber\\
    &-\sum_{i}^{z'} K_{1,i}^{*}b_{1,i}\sigma_{1}-\sum_{i}^{z'} K_{2,i}^{*}b_{2,i}\sigma_{2}\label{h22} \ ,
\end{align}
where \( z' \) is the number of second neighbors, and \( K^{*}_{1,i} := J^{*}_{1,i} / (k_B T) \) represents the rescaled coupling energy for second neighbors. Similar to the nearest-neighbor case, the spin magnetization for the 1 and 2 spin Hamiltonians is calculated, resulting in the following expressions 
 
\begin{align}
    \langle \sigma_{1}\rangle_1=&\sum_{i}^{z} K_{1,i}'b_{1,i}'+\sum_{i}^{z'} K_{1,i}'^{*}b_{1,i}' \ ,\label{magne21}\\
    \langle \sigma_{1}\rangle_1=&\sum_{i}^{z-1} K_{1,i}b_{1,i}+\sum_{i}^{z'} K_{1,i}^{*}b_{1,i} \nonumber\\
    +&\tanh(K_{1,2})\Bigg(\sum_{i}^{z-1} K_{2,i}b_{2,i}+\sum_{i}^{z'} K_{2,i}^{*}b_{2,i}\Bigg) .\label{magne22}
\end{align}
Subsequently, to perform the configurational averaging as in the case of nearest neighbors, it is necessary to modify the bond probability density already established in equation \ref{modeloFeAl}, adding the second neighbor interaction, taking into account the configuration in which this interaction occurs. For this investigation, it is proposed to work with a probability density of the form:
\begin{align}
    & P(K_{i,j})=\overbrace{p^{2}\delta(K_{i,j}-K_{3}(p))+(1-p^2)\delta(K_{i,j})}^{\textnormal{1 neighbors}}+ \nonumber\\
    &\overbrace{p^2(1-p)\delta(K^{*}_{i,j}-K_4(p))+(p^{3}+1-p^{2})\delta(K^{*}_{i,j})}^{\textnormal{2 neighbors} } \label{modeloFeAl2},  
\end{align} 
where, as can be observed, the first two terms pertain to the well-established density of nearest neighbors, the third term represents the second-neighbor interaction of the  Fe-Al-Fe  type, and the last term accounts for other second-neighbor configurations that do not exhibit any interaction.

By incorporating the revised bond probability density and utilizing the magnetization expressions (Equations \ref{magne21} and \ref{magne22}), the configurational average (Equation \ref{sigma_prome}), and the renormalization equation provided by the MFRG method (Equation \ref{EcuaRenorma}), we can, after detailed calculations, derive an equation that accurately models the \( Fe_pAl_{1-p} \) system near the transition point while accounting for interactions with second-nearest neighbors. This result is shown 
\begin{align}
   0 =& -1 + \frac{p^2 \tanh\Big(\frac{J_{4}(P)}{k_B T}\Big)}{J_{3}(p)}   \nonumber  \\
   & \Big[(z-1)J_{3}(p)+J_{4}(p)(1-p)z'\Big], \label{vecinos2}
\end{align}
which similiarly as we use $\mathcal{E}(J_{3}(p),J_{4}(p))$ to denote the righ hand side of the equation, such that:
\begin{equation}\label{eq3}
    \mathcal{E}(J_3(p),J_{4}(p))=0
\end{equation}

\section{THE NEURAL NETWORK APPROACH FOR FITTING THE DATA WITH THE MODEL}\label{sec:neural_network_approach}
Neural networks have become increasingly prevalent for developing numerical solution methods aimed at solving equations, particularly partial differential equations, across various research domains. Their rising popularity can be attributed to the numerous advantages they offer over conventional methods. For example, neural networks are particularly adept at approximating highly non-linear and complex functions, enabling them to capture intricate relationships among variables.

Following the definition provided by \cite{calin2020deep}, a neural network can be characterized as a computational model comprised of interconnected nodes, commonly referred to as neurons or artificial neurons, organized into layers. Each neuron in the network is associated with an \textit{activation function} \( \varphi \). This activation function introduces non-linearity into the model, allowing it to learn complex patterns. The output of a neuron \( f \), after applying the activation function, can be expressed as a weighted sum of inputs plus a bias term, followed by the activation function:

\begin{equation*}
    f =  \varphi \left( \sum_{i} \omega_{i} y_{i} + b \right),
\end{equation*}
where the set of \( \omega_i \) represents the weights of the neuron, \( y_i \) denotes the inputs, and \( b \) is the bias term. The layers, on the other hand, serve as structural components that organize all the neurons into groups. There are three primary types of layers: input, hidden, and output layers. The input layer corresponds to the input data of the neural network, with each node in this layer representing a feature or attribute of the input. The hidden layers consist of various ensembles of neurons, each equipped with its activation functions and weights. Finally, the output layer generates the network's output or prediction.

Consequently, we can define a learning function \( \mathcal{F} \) as the function composed of all the neurons in our neural network, which depends on the complete set of parameters \( \{\omega\} \), \( \{b\} \), and activation functions. In our particular case, we will define our learning function as follows
 \begin{equation}\label{learningfunction}
\mathcal{F}  : \mathbb{R}^2 \rightarrow \mathbb{R}^n, \qquad
(p, T) \mapsto (J_0, .... J_{n}),
\end{equation}
where \( p \) and \( T \) serve as the inputs, or data, that feed our neural network for each case under consideration, while \( J_i \) represents the respective coupling energy functions.

Training the neural network involves adjusting its parameters to minimize a specific function, commonly referred to as the \textit{cost function}. This function is selected based on the nature of the task at hand (e.g., binary classification, multi-class classification, regression, etc.). For our particular case, in accordance with the work of \cite{blechschmidt2021three}, we propose the following cost function 
\begin{equation}\label{cost_function}
    C(\mathcal{F}) =\frac{1}{N}\sum^{N}_i \lVert  \mathcal{E}( \mathcal{F} )|_{i} \rVert,
\end{equation}

where \( \mathcal{E}(\mathcal{F})|_{i} \) represents the right-hand side of the state equations for the case under consideration, namely equations \ref{eq1}, \ref{eq2}, and \ref{eq3}, evaluated at the experimental points \( (p_i, T_i) \) for \( i = 1, \dots, N \) in each case.

From the above, our primary objective is to approximate the solutions of the equation of state for each configuration by identifying the optimal combination of parameters and activation functions that minimize \( C(\mathcal{F}) \) such that \( C(\mathcal{F}) \approx 0 \). To achieve this, we treat the cost function \( C(\mathcal{F}) \)  as a function of the parameters within the learning function \( \mathcal{F} \). Subsequently, we can employ a standard optimization algorithm, such as the gradient descent method \cite{yadav2015introduction}, to search for the minimum.

\section{NUMERICAL RESULTS}\label{sec:numerical_results}
Since the neural network approach for fitting experimental data involves optimizing the parameters of the learning function during training\ref{learningfunction}, the fitting process is inherently non-unique. Given that neural networks start with random initial conditions, it is crucial to evaluate the model's stability and reliability. To address this, in this work, the fitting process was repeated multiple times. This allowed us to assess the consistency of the results and ensure that the model is not overly sensitive to specific starting points. The results are depicted in graphs, where a solid line represents the average of the fitted data, and a shaded area indicates the total variability.

\subsection{\texorpdfstring{\( Fe_pAl_{1-p} \)} . system}

\begin{figure}[htbp]
    \centering
    \includegraphics[width=1\linewidth]{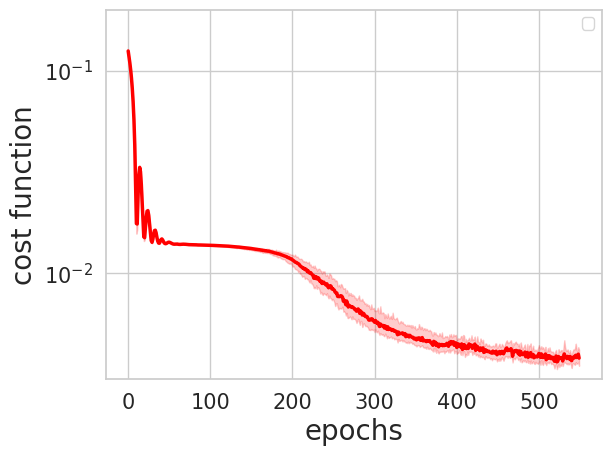}
    \caption{Minimization of the error function used in the neural network fit for the \( Fe_pAl_{1-p} \) system, where highlighting the mean of all attempts in a dark line .}
    \label{FeAl_error_1veci}
\end{figure}

We begin by analyzing the \( Fe_pAl_{1-p} \) system, considering only first-neighbor interactions and substituting the expression in Equation \ref{eq2} into the cost function \ref{cost_function}. For this system, after training the neural network, a significant reduction in the cost function was achieved, reaching an order of magnitude of \( 10^{-2} \) and approaching \( 10^{-3} \), as illustrated in Figure \ref{FeAl_error_1veci}. This result provides strong evidence supporting the validity of both the model and the neural network approach. The fitting process employed a neural network that combined various activation functions, including \textit{tanh}, \textit{softmax}, \textit{selu}, and \textit{gelu}. For a more detailed discussion of these activation functions, see \cite{calin2020deep}.

Following this, we obtained the behavior of the Fe-Fe bonding energy for this system, as shown in Figure \ref{FeAl_energy_1veci}. Comparing the results obtained using a neural network approach with those derived from linear models, as in recent studies \cite{Blume-Capel2,rodriguez2023}, it is evident that the neural network model provides greater generality and imposes fewer constraints on the system. Moreover, considering that the concentration in the system is proportional to the interatomic distance, as proposed in previous studies \cite{distan_inter_atomic1,distan_inter_atomic2}, we observe that Figure \ref{FeAl_energy_1veci} resembles a plot of interaction energy versus interatomic distance. This behavior aligns closely with the Bethe-Slater curve \cite{Beth-slater}, which characterizes the bonding energy of metals as a function of interatomic distance, including that of iron.

Additionally, by examining the quantitative values of the energy produced by the fit, we observe good agreement with expected values, as the bonding energy of ferromagnetic metals such as iron and nickel typically lies within the range of \( 10-50 \, \mathrm{eV} \) \cite{Enlace_pares_segunda,transiciones}. Notably, without imposing any constraints on the function fit, positive bonding energies were obtained, consistent with the expected behavior for Fe-Fe type bonds.

Furthermore, it is worth noting that, although the bond energies of these systems diverge at certain points, the phase diagram generally shows minimal divergence. Across all iterations, the values remain close to the average, providing additional support for the convergence not only of the minimization method employed in the network but also of the MFRG method.

\begin{figure}[htbp]
    \centering
    \includegraphics[width=1\linewidth]{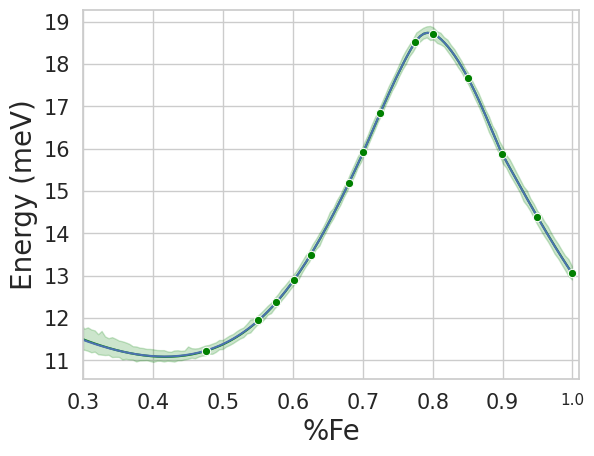}
    \caption{Graph of the Fe-Fe binding energy, obtained through neural network fitting, where the points correspond to the energy values at the locations presented by the experimental data in Figure \ref{FeAl_exp_dat} and highlighting the mean of all attempts in a dark line.}
    \label{FeAl_energ_1veci}
\end{figure}

\begin{figure}[htbp]
    \centering
    \includegraphics[width=1\linewidth]{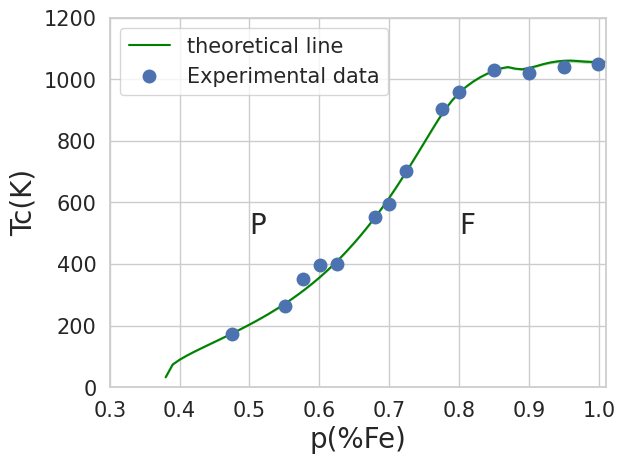}
    \caption{Magnetic phase diagram of the $Fe_pAl_{1-p}$ system, obtained through neural network fitting, where highlighting the mean of all attempts in a dark line.}
    \label{FeAl_diagram_1veci}
\end{figure}

Using the energy obtained from the previously presented fit and plotting the expression given by equation \ref{expModeFeAl}, we obtained the phase diagram of the \( Fe_pAl_{1-p} \) system, as shown in Figure \ref{FeAl_diagram_1veci}. The close alignment of the data points with the theoretical line is notable, as well as the characteristic percolation point of this system, marked by a sharp drop in critical temperature at low iron concentrations, where the presence of Fe-Fe bonds is minimal.

Finally, the last figures presented show how the properties of the data are preserved in the average of each graph. This demonstrates that the model is valid and not merely dependent on a local minimum of the error function.

\subsection{\texorpdfstring{$Fe_pMn_{0.6-p}Al_{0.4}$ }  .system}

\begin{figure}[htbp]
    \centering
    \includegraphics[width=1\linewidth]{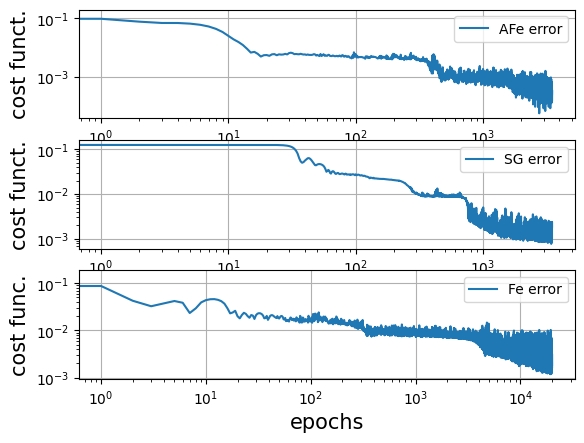}
    \caption{Graphs showing the behavior of the error functions used in the neural network to fit expressions \ref{expModeFeMnAl} and \ref{expModeFeMnAlSg}. The first corresponds to the antiferromagnetic phase, the second to the spin glass phase, and the third to the ferromagnetic phase.}
    \label{FeMnAl_error}
\end{figure}

Turning to the \( Fe_pMn_{0.6-p}Al_{0.4} \) system, we observe three distinct types of transitions, each reflected in separate adjustments. By following a procedure similar to the one used for the previous system, for each transition, and incorporating the expression \ref{eq1} into the cost function \ref{cost_function}, we obtain three graphs for the error function corresponding to each transition. These graphs are consistent with those observed previously, showing an error close to \( 10^{-3} \) in the order of \( 10^{-2} \) across all adjustment phases, as shown in Figure \ref{FeMnAl_error}.

\begin{figure}[htbp]
    \centering
    \includegraphics[width=1\linewidth]{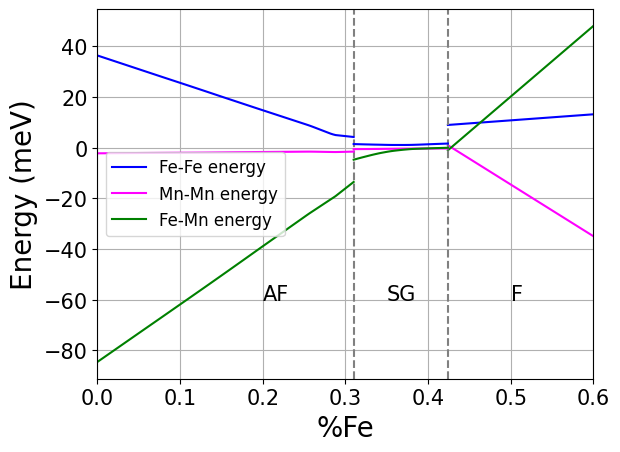}
    \caption{Graphs showing the behavior of the fitted binding energies in the different magnetic phases, where the blue line corresponds to the energy of the Fe-Fe bonds, the magenta line to the energy of the manganese-manganese bonds, and the green line to the iron-manganese bonds. }
    \label{FeMnAl_energy}
\end{figure}

From the results above, the behavior of the bond energy fit is shown in Figure \ref{FeMnAl_energy}. Despite offering greater flexibility in the energy model compared to previous studies, the common issue of discontinuities at phase transition points persists. However, unlike the results reported in previous work \cite{rodriguez2023}, this model manages to reduce the energy discontinuities to less than 20 MeV.

Additionally, Figure \ref{FeMnAl_energy} clearly shows that the signs of the energies are largely consistent with expectations. Specifically, the Fe-Fe bond energy is ferromagnetic (positive), the Mn-Mn bond energy is antiferromagnetic (negative), and the Fe-Mn bond energy is generally antiferromagnetic. However, a shift toward the ferromagnetic side is observed in the final transition for the Fe-Mn bonds, which may be due to the influence of second-neighbor interactions not considered in this model. Furthermore, the bonding energy values for metals, such as pure iron, fall within the expected range. Near its maximum concentration, the energy value aligns with the anticipated range, providing further validation for the proposed model.

\begin{figure}[htbp]
    \centering
    \includegraphics[width=1\linewidth]{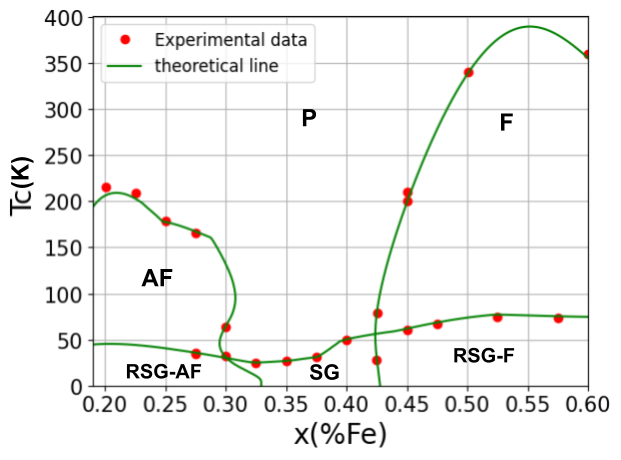}
    \caption{Graph of the phase diagram of the \( Fe_pMn_{0.6-p}Al_{0.4} \) system obtained through fitting using neural networks.}
    \label{FeMnAl_diagram}
\end{figure}

Finally, by incorporating the adjusted energy functions into equations \ref{expModeFeMnAl} and \ref{expModeFeMnAlSg}, the fit shown in Figure \ref{FeMnAl_diagram} is obtained. This figure demonstrates a good fit to the data, highlighting the presence of percolation points in the iron transition. It is important to note that, in contrast to the previously shown data graph, this fit excludes data related to superparamagnetic transitions. This exclusion is due to the fact that the model used in this study does not account for magnetic phases induced by nanoparticles or nanodomains, which are the primary cause of superparamagnetism.

\subsection{\texorpdfstring{$Fe_pAl_{1-p}$}. system with second-neighbor interactions}

\begin{figure}[htbp]
    \centering
    \includegraphics[width=1\linewidth]{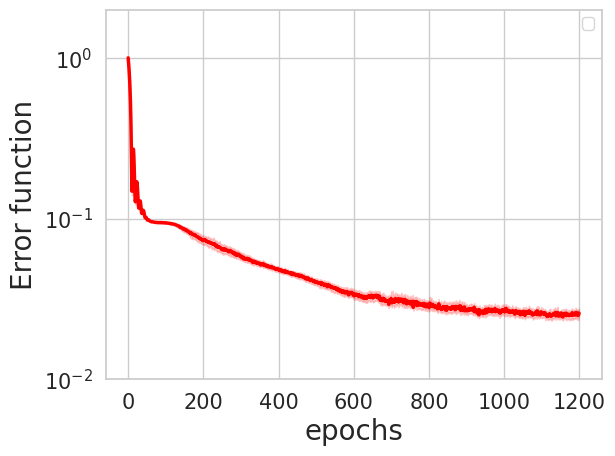}
    \caption{Graph showing the behavior of the error function for the \( Fe_pAl_{1-p} \) system when implementing second neighbors. The shaded area represents the range of all possible values taken in all attempts, while the solid line represents the mean value.}
    \label{FeAl_error_2veci_muchos}
\end{figure}

In this case, by incorporating the expression \ref{eq3} into the cost function, the same process used for previous systems can be applied with a neural network fit. Similar to the fitting performed for first neighbors, this fit was repeated multiple times to ensure the reliability of the method and to confirm that the result is not a local minimum of the error function. Additionally, as with the first-neighbor case, a reliability range was established for the fit, which is represented in the graphs as the range of all possible outcomes. The first result obtained for this system is the behavior of the error function during the fit, shown in Figure \ref{FeAl_error_2veci_muchos}, where a decreasing trend of the order of $10^{-1}$ is observed. While this is not an exceptionally low number, it represents a significantly small error compared to the values being fitted, which are on the order of $10^{-3}$.

\begin{figure}[htbp]
    \centering
    \includegraphics[width=1\linewidth]{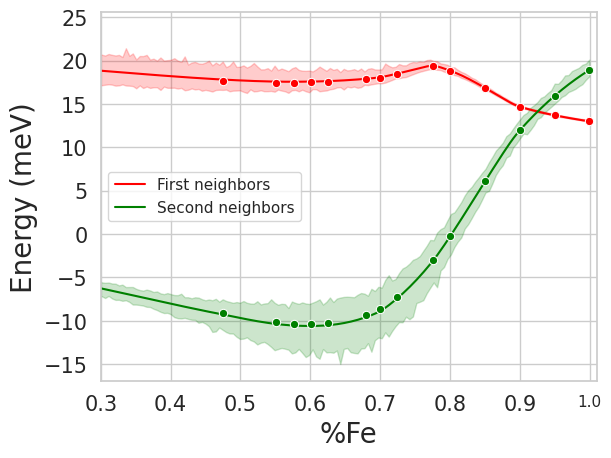}
    \caption{Graph showing the behavior of the bond energy for second and first neighbors, in the red and green graphs respectively, with the shaded areas representing all possible values taken in all attempts and the solid lines representing the mean value.}
    \label{FeAl_energy_2veci_muchos_dat}
\end{figure}

In a manner analogous to the previously studied systems, the behaviors of the bonding energies were also obtained. In this case, the reference is to the bonding energies of first and second neighbors, which are shown in Figure \ref{FeAl_energy_2veci_muchos_dat}.

Observing the graph shown in Figure \ref{FeAl_energy_2veci_muchos_dat}, the signs of the energies stand out. This is particularly important for the second-neighbor energy because its presence is limited to points where the proportions of iron and aluminum are similar, which promotes the generation of these bonds. This is reflected in the aforementioned graph, as the energy is generally antiferromagnetic and becomes more pronounced at points of similar concentration. Additionally, the first-neighbor energy of iron retains its characteristic sign (ferromagnetic) and remains within the previously mentioned range.

Subsequently, revisiting the previously presented data, we obtain the fit shown in Figure \ref{FeAl_energy_2veci_muchos_dat}. It can be observed that, although there are points in the energy where the data deviate from the mean, these deviations are minimal and nearly imperceptible within the adjustment. As a result, the model maintains a good fit, as shown in \ref{FeAl_diagram_1veci}, while also incorporating the interaction with second neighbors, which had been previously neglected in other studies. 

\begin{figure}[htbp]
    \centering
    \includegraphics[width=1\linewidth]{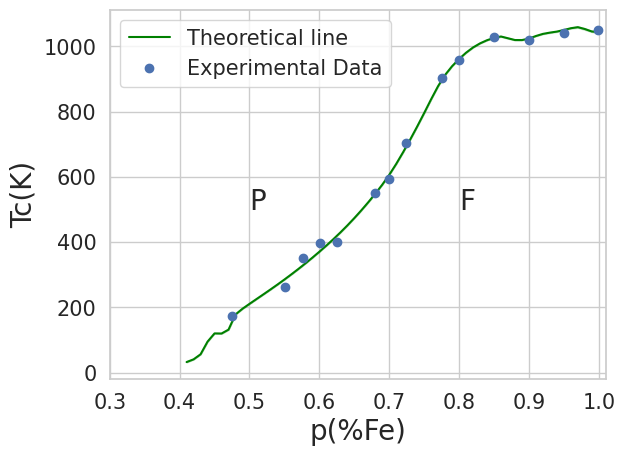}
    \caption{Graph of the phase diagram obtained by fitting with neural networks, implementing second neighbors in the $Fe_pAl_{1-p}$ system.}
    \label{FeAl_diagram_2veci_muchos_dat}
\end{figure}

We conclude this section by noting that the flexibility offered by neural networks in modeling bond energy functions leads to excellent fitting results. These fits not only exhibit high accuracy but also yield energy behaviors that align with both quantitative and qualitative physical expectations. Additionally, key features such as percolation points and phase transitions are clearly identified in the resulting diagrams. 

\section{Conclusion}\label{discussion}

In previous studies employing similar methods, predefined behaviors were often imposed on the binding energy during the fitting process, typically using constant \cite{Perez1,FeNiMn}, linear \cite{Perez2,rodriguez2023}, or cubic \cite{cubica} models. In contrast, this work utilizes a neural network approach to derive the binding energy function, eliminating the need for prior assumptions about its form. This represents a significant advancement over traditional models, as it offers greater flexibility in fitting the binding energy behavior, enabling it to more accurately reflect experimental data and align with physical reality. Furthermore, this method facilitates extrapolation beyond the experimental range—a capability that simple interpolation techniques lack due to their inherent limitations. The flexibility of the neural network approach not only ensures a more precise fit to experimental data but also enhances the model's ability to capture the underlying physics. Beyond improving known models and reproducing previous results, this approach expands the potential for extending the analysis to more complex systems, such as those incorporating second-neighbor interactions, thus broadening the scope for future research.

\nocite{*}
\bibliography{Article} 

\end{document}